\documentclass[11pt,a4paper]{article}

\usepackage{epsfig}
\usepackage{amsmath}
\usepackage{amssymb}
\usepackage{slashed}
\usepackage{verbatim}
\usepackage{bm}
\usepackage{dsfont}
\usepackage[footnotesize]{caption}
\usepackage{subfigure}
\usepackage{cancel}
\usepackage{cite}
\usepackage{hyperref}
\usepackage{xcolor}
\usepackage{hyperref}
\hypersetup{
colorlinks = true,
linkcolor=blue,
citecolor=green!40!black,
urlcolor=green!40!black,
}

\usepackage{multirow}
\usepackage{bbold}

\definecolor{rossoferrari}{HTML}{D9073D}
\definecolor{mediumblue}{HTML}{0000CD}
\definecolor{forestgreen}{HTML}{228B22}
\definecolor{desy_blue}{HTML}{009EE2}
\definecolor{desy_orange}{HTML}{FD8800}
\definecolor{light_pink}{rgb}{1,0.4,0.4}
\definecolor{light_blue}{rgb}{0.284602,0.317763,0.963947}

\title{\LARGE\bf Discovery Opportunities with Gravitational Waves \\ \vspace*{5mm} TASI 2024 Lecture Notes}
\author{Valerie Domcke}
\date{\textit{\small Theoretical Physics Department, CERN, 1 Esplanade des Particules, Geneva, Switzerland}}

\begin{document}

\maketitle

\vskip 1cm

\begin{abstract}
 Recent advancements in gravitational wave astronomy hold the promise of a completely new way to explore our Universe. These lecture notes aim to provide a concise but self-contained introduction to key concepts of gravitational wave physics, with a focus on the opportunities to explore fundamental physics in transient gravitational wave signals and stochastic gravitational wave background searches.
\end{abstract}

\newpage
\tableofcontents

\newpage
\section{Lecture I - Introduction to Gravitational Waves}
\label{sec:lect1}

Most of what we know about our Universe and a lot of what we know about particle physics stems from precision observations in a wide range of electromagnetic frequencies. However, these observations come with a fundamental limit on how far `back in time' we can observe our universe, or equivalently, a limit on the energy scale of the particle physics involved: at times before CMB decoupling, when the universe was filled with charged particles, it was not transparent to electromagnetic radiation, making it very difficult to obtain any information on this early epoch through electromagnetic radiation. To pierce this barrier, we seek a messenger which interacts more weakly with the matter content of our universe. This is one of promises of gravitational wave (GW) astronomy: GWs can traverse our Universe essentially unperturbed, providing information both about their source and on the expansion history of the universe - if(!) we can detect them. The weakness of gravity is thus both a key advantage and disadvantage at once : we can probe deeper into our cosmic history, but the detection is correspondingly more challenging.

This first direct detection of gravitational waves was in 2015, when the LIGO/Virgo collaboration detected the first merger of two black holes at a distance of about 400~Mpc (1.3 billion light years)~\cite{LIGOScientific:2016aoc}. In terms of cosmic history, this is considered a fairly recent event, and has not yet demonstrated to unlock the potential eluded to above. However, data is rapidly accumulating and new detectors are under construction. In these lectures, I will review the current status and prospects of this rapidly growing field. Lectures~\ref{sec:lect1} and~\ref{sec:lect2} will introduce key concepts of gravitational wave physics, allowing us to describe the sourcing, propagation and detection of gravitational waves. Lecture~\ref{sec:lect3} will focus on discovery opportunities for fundamental physics from transient signals (in particular black hole and neutron star mergers) while Lecture~\ref{sec:lect4} will lead us to truly early universe sources and their imprint on the stochastic GW background.

\paragraph{Literature.} Large parts of lectures~\ref{sec:lect1} and~\ref{sec:lect2} follow Ref.~\cite{Maggiore:2007ulw}. For a detailed review on stochastic GW backgrounds, see e.g.~\cite{Caprini:2018mtu}.

\subsection{Wave equation.} Our starting point is Einstein's equation,
\begin{align}
 G_{\mu\nu} \equiv R_{\mu\nu} - \frac{1}{2} g_{\mu\nu} R = 8 \pi G T_{\mu \nu} \,,
 \label{eq:einstein}
\end{align}
with $R$ denoting the Ricci scalar, $R_{\mu\nu}$ the Ricci tensor, $g_{\mu\nu}$ the metric tensor, $G$ Newton's constant and $T_{\mu\nu}$ the energy momentum tensor. The left-hand side of this equation is fully determined by the space-time metric $g_{\mu\nu}$, the right-hand side describes the matter content of the universe. In a nutshell, this equation relates the two: matter curves the space-time metric, and vice versa, the metric determines the geodesic of any test particles.

We consider a small departure from flat spacetime, $g_{\mu\nu} = \eta_{\mu\nu} + h_{\mu\nu}(x)$, where $\eta_{\mu\nu}$ is the flat Minkowski metric and $h_{\mu\nu}$ with all entries $|h_{\mu\nu}| \ll 1$ is a small perturbation which we will identify as a GW.\footnote{We will keep the order in $h$ explicit, allowing us to raise and lower indices with the flat metric $\eta_{\mu\nu}$.} This assumption significantly simplifies the left-hand side of Eq.~\eqref{eq:einstein}, which we will now compute to linear order in $h$. We start with the Christoffel symbol,
\begin{align}
  \Gamma^\mu_{\nu\beta} &\equiv \frac{1}{2} g^{\nu \sigma} \left[ \partial_\nu g_{\beta \sigma} + \partial_\beta g_{\nu\sigma} - \partial_\sigma g_{\nu \beta} \right] \nonumber \\
  & = \frac{1}{2} \eta^{\nu \sigma} \left[ \partial_\nu h_{\beta \sigma} + \partial_\beta h_{\nu\sigma} - \partial_\sigma h_{\nu \beta} \right] + {\cal O}(h^2)\,,
\end{align}
where we have used that $h$, but not $\eta$, is allowed to be space-time dependent. From this, we obtain the Riemann curvature tensor as
\begin{align}
 R^\mu_{\nu\alpha\beta} & \equiv \partial_\alpha \Gamma^\mu_{\nu \beta} - \partial_\beta \Gamma^\mu_{\nu \alpha} + \Gamma^\mu_{\alpha \sigma} \Gamma^\sigma_{\nu \beta} - \Gamma^\mu_{\beta \sigma} \Gamma^\sigma_{\nu\alpha} \nonumber \\
 & = \frac{1}{2} \left[ \eta^{\mu \sigma} (\partial_\alpha \partial_\nu h_{\beta \sigma} + \partial_\alpha \partial_\beta h_{\nu \beta} - \partial_\alpha \partial_\sigma h_{\nu \beta}) - (\alpha \leftrightarrow \beta) \right] + {\cal O}(h^2)
\end{align}
Here, we have dropped the last two terms in the first line since $\Gamma = {\cal O}(h)$. The second term in the second line is symmetric under exchanging $\alpha$ and $\beta$, and hence will cancel with the corresponding term stemming from the second term in the first line. Contracting $R^\alpha_{\mu \alpha \nu} = R_{\mu \nu}$ and $g^{\mu\nu} R_{\mu\nu} = R$ we obtain (after some algebra) for the left-hand side of the linearized Einstein equation,
\begin{align}
 G_{\mu\nu} = - \frac{1}{2} \left[ \Box \bar h_{\mu\nu} + \eta_{\mu\nu} \partial^\rho \partial^\sigma \bar h_{\rho\sigma} - \partial^\rho \partial_\nu \bar h_{\mu\nu} - \partial^\rho \partial_\mu \bar h_{\nu \rho} \right]
 \label{eq:Gmunu}
\end{align}
where for notational convenience we have introduced
\begin{align}
 \bar h_{\mu\nu} \equiv h_{\mu\nu} - \frac{1}{2} \eta_{\mu\nu} h \,, \quad h \equiv h^\mu_\mu = \eta_{\mu \nu} h^{\nu \mu} \,.
\end{align}

We can simplify this expression further by fixing a gauge, which in general relativity (GR) corresponds to fixing a frame for the observer. By considering an infinitesimal coordinate transformation on all four coordinates, $x^\mu \mapsto x^\mu + \epsilon^\mu$ and $h_{\mu\nu} \mapsto h_{\mu\nu} - (\partial_\mu \epsilon_\nu + \partial_\nu \epsilon_\mu)$, we can gauge fix four degrees of freedom out of the 10 degrees of freedom contained in the symmetric tensor $h_{\mu\nu}$. Analogous to electromagnetism, one can use this to satisfy the Lorenz gauge condition,
\begin{align}
 \partial_\mu \bar h^{\mu\nu} = 0\,.
\end{align}
One can immediately see that this significantly signifies Eq.~\eqref{eq:Gmunu}, so that in Lorenz gauge, we obtain the linearized Einstein equation
\begin{align}
 \Box \bar h_{\mu\nu} = - 16 \pi G T_{\mu\nu}^\text{(an)} \,.
 \label{eq:wave_equation}
\end{align}
The box operator on the left-hand side marks this as a wave equation, the energy momentum tensor on the right-hand side can be viewed as a source term. Here the label 'an' for anisotropic indicates that we have subtracted the background energy momentum tensor which solves Einstein's equation to 0th order in $h$. This equation fully describes the sourcing and propagation of GWs.

At this point we note that we still have a residual gauge freedom since the Lorenz gauge condition allows to perform a further coordinate transformation $\tilde \epsilon^\mu$ with $\Box \tilde \epsilon^\mu = 0$. This permits us to fix another four degrees of freedom, leaving us with a remaining two degrees of freedom for the GW. In vacuum, $T_{\mu\nu} = 0$ this can be conveniently done by choosing the transverse traceless (TT) gauge, $\bar h = 0$ and $\bar h_{0i} = 0$, which amounts to four gauge conditions. Combined with Lorenz gauge and Eq.~\eqref{eq:wave_equation} with $T_{\mu\nu}=0$, this gives the TT gauge conditions
\begin{align}
 h_{0\mu}^{TT} = 0 \,, \quad h^{TT} = 0 \,, \quad \partial^j h_{ij}^{TT} = 0 \,,
\end{align}
which significantly simplify the treatment of GWs.
In the vicinity of a source $T_{\mu\nu} \neq 0$, the same residual gauge freedom exists but conditions for gauge fixing take a more complicated form. We will discuss the physical meaning of the TT gauge below.

In summary, we have seen that the gravitational wave contains two degrees of freedom, obeys a wave equation, travels at the speed of light in vacuum and is sourced by the anisotropic part of the energy momentum tensor.

\subsection{Expanding FRW universe}
So far, we have taken the background metric to be Minkowski. To be able to describe GWs in cosmological setups, we generalize this to an expanding Friedmann-Lemaitre-Walker (FRW) universe expanding with a scale factor $a(t)$,
\begin{align}
 ds^2 = g_{\mu\nu} x^\mu x^\nu = a^2(\tau) (\eta_{\mu\nu} + h_{\mu\nu}) x^\mu x^\nu \rightarrow a^2(\tau) \left(- d\tau^2 + (\delta_{ij} + h^{TT}_{ij}) dx^i dx^j \right)\,.
 \label{eq:FRWmetric}
\end{align}
Here we have introduced co-moving coordinates $x^\mu$, including conformal time $x^0 = \tau$ with $dt = a \, d\tau$. In the following, spatial and covariant derivatives are taken with respect to the co-moving coordinates, and for time derivatives we distinguish with a prime the derivative with respect to conformal time and with dot the derivative with respect to cosmic time. The last step in Eq.~\eqref{eq:FRWmetric} holds in TT gauge.

Following the same steps as above, we obtain after somewhat lengthy algebra\footnote{
Note that the left-hand side is just the Klein-Gordon equation in FRW space time,
\begin{align}
 \frac{1}{\sqrt{-g}} \partial_\mu (\sqrt{-g} g^{\mu\nu} \partial_\nu) \phi = a^{-2} (\Box \phi - 2 \frac{a'}{a} \phi') = 0 \,, \nonumber
\end{align}
where $\sqrt{-g} \equiv \sqrt{- \text{det}(g_{\mu\nu})} = a^4$ and $g_{\mu\nu} = \text{diag}(- a^2, a^2, a^2, a^2)$ and $g^{\mu\nu}$ denoting the inverse metric.
}
\begin{align}
   \Box \bar h_{\mu\nu} - 2 \frac{a'}{a} \bar h'_{\mu\nu} = - a^{-1} \left( \tilde h''_{\mu\nu} - (\vec \nabla^2 + \frac{a''}{a}) \tilde h_{\mu\nu} \right) =  16 \pi G T_{\mu\nu}^{(an)} \,,
 \label{eq:wave_equation_FRW}
\end{align}
with $\Box = \eta^{\mu \nu} \partial_\mu \partial_\nu = - \partial_\tau^2 + \vec \nabla^2$ and $\tilde h_{\mu\nu} \equiv a \bar h_{\mu\nu}$. In a static universe, $a' = 0$ and we trivially recover Eq.~\eqref{eq:wave_equation}.
Let us analyze this equation in vacuum, choosing TT gauge and performing a Fourier transformation,
\begin{align}
  \tilde h_{ij}^{''}(\vec k, \tau) + (k^2 - \frac{a''}{a}) \tilde h_{ij}(\vec k, \tau) =  0\,,
  \label{eq:we_FRW_fourier}
\end{align}
where $\vec k$ denotes the co-moving wave vector. Identifying $a''/a = (aH)^2$ where $H = \dot a /a$ is the Hubble parameter, we distinguish two distinct cases.
\begin{itemize}
 \item For $k \gg a H$ we are in the sub-horizon regime, i.e.\ the GW wavelength is small compared to the Hubble horizon. Eq.~\eqref{eq:we_FRW_fourier} simplifies to a standard wave equation for $\tilde h$,
 \begin{align}
  \tilde h''_{ij} + k^2 h_{ij} = 0
 \end{align}
 and hence for $h_{ij} = \tilde h_{ij}/a$ we obtain a wave with an amplitude that decays as $1/a$.
 \item For $k\ll aH$ we are in the super-horizon regime, i.e.\ the GW wavelength is large compared to the Hubble horizon. To analyze this, we express Eq.~\eqref{eq:we_FRW_fourier} in terms of $h_{ij}$, which yields
 \begin{align}
  2 a' h'_{ij} + a h''_{ij} = 0 \,.
 \end{align}
This is formally solved by $h_{ij} = A_{ij} + B_{ij} \int_0^\tau \frac{d\tau'}{a^2(\tau')}$ with $A$ and $B$ constants of integration. In an expanding universe, the second term decreases with time, so that eventually, the GW is simply given by the integration constant $A_{ij}$, which is obtained from matching the boundary conditions at the point in time when the GW crossed the Hubble horizon. Note that the `gravitational wave' is simply a constant on super-horizon scales, which is often referred to as `freezing' of GWs outside the Hubble horizon.

\end{itemize}

\subsection{Popular choices of frames}
To understand the physical meaning of popular gauge choices, such as the TT gauge above, let us consider the effect of a GW on two test masses located at $x^\mu$ and $x^\mu + \xi^\mu$. For simplicity, let us take the background metric to be Minkowski. From the geodesic equation,
\begin{align}
 \frac{d^2 x^\mu}{d\tau^2} + \Gamma^\mu_{\nu\rho}(x) \frac{dx^\nu}{d\tau} \frac{dx^\rho}{d\tau} = 0\,,
 \label{eq:geodesic}
\end{align}
evaluated at $x^\mu$ and $x^\mu + \xi^\mu$, we obtain the geodesic deviation equation describing the evolution of the relative distance $\xi^\mu$ between these two points,
\begin{align}
 \frac{d^2 \xi^\mu}{d\tau^2} + 2 \Gamma^\mu_{\nu \rho}(x)  \frac{dx^\nu}{d\tau} \frac{d\xi^\rho}{d\tau} + \xi^\sigma \partial_\sigma \Gamma^\mu_{\nu \rho}(x)  \frac{dx^\nu}{d\tau} \frac{dx^\rho}{d\tau} + {\cal O}(\xi^2) = 0\,.
 \label{eq:geodesic_dev}
\end{align}

\paragraph{TT frame.} Consider a test mass initially at rest, $dx^i/d\tau = 0$ at $\tau = \tau_0$. Inserting this into Eq.~\eqref{eq:geodesic} gives at $\tau = \tau_0$,
\begin{align}
 \frac{d^2 x^i}{d\tau^2}  = -  \Gamma^i_{00}(x) \left(\frac{dx^0}{d\tau} \right)^2 = 0\,.
\end{align}
Explicitly writing $\Gamma^i_{00} = (2 \partial_0 h_{0i} - \partial_i h_{00})/2$ we see that this vanishes in the TT gauge. Thus test masses initially at rest stay at rest in this frame, or in other words, the coordinates of this frame are set by freely falling test masses. The TT gauge has the benefits of a very simple structure for the GW and a clear physical interpretation.

\paragraph{Proper detector frame.} The proper detector frame, or laboratory frame, is constructed using Fermi normal coordinates. We choose a gauge in which the metric is locally flat, $\Gamma^\mu_{\nu\rho}(x_0) = 0$, with $x_0$ a reference point in the laboratory. We construct a local inertial frame centered around this point, with a coordinate system defined by hypothetical rigid rulers (i.e.\ which are not impacted by the GW). Assuming a non-relativistic detector, $dx^i/d\tau \ll dx^0/d\tau = 1$, Eq.~\eqref{eq:geodesic_dev} applied to $x_0^\mu$ and $x_0^\mu + \xi^\mu$ becomes
\begin{align}
 \ddot \xi^i = - \xi^\sigma \partial_\sigma \Gamma^i_{0 0}(x_0) = R^i_{0 j 0} \xi^j \,.
\end{align}
In the second step we have used $\partial_0 \Gamma^i_{00}= 0$, since the metric in the proper detector frame takes the shape $g_{\mu\nu} = \eta_{\mu\nu} + {\cal O}((x^i - x_0^i)( x^j-x_0^j))$ and thus non-zero contributions at $\vec x_0$ arise only for two spatial derivatives acting on $g_{\mu\nu}$.
Since the Riemann tensor is gauge invariant at linear order in $h$, we can choose to evaluate it in TT gauge,
\begin{align}
 R^i_{0j0} = - \frac{1}{2} \ddot h_{ij}^{TT}\,,
\end{align}
to obtain
\begin{align}
 \ddot \xi^i = \frac{1}{2} \ddot h_{ij}^{TT} \xi^j\,.
 \label{eq:xi_eom}
\end{align}
The right-hand side can be interpreted as a Newtonian force acting on the test particles, yielding a clear physical interpretation of the effect of the GW in this frame.

\begin{figure}
\centering
 \includegraphics[width = 0.8 \textwidth]{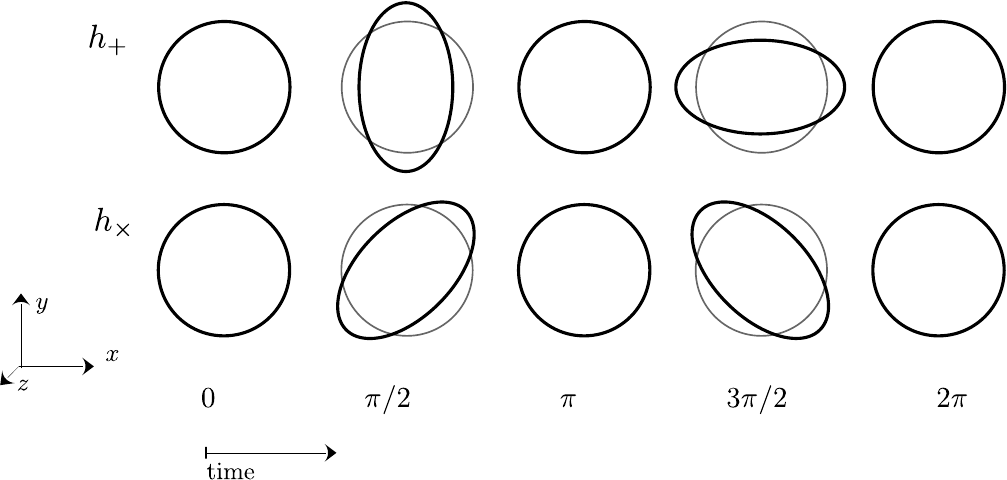}
 \caption{The two GW polarizations. Motion of a circle of test particles in a plane orthogonal to the GW propagation direction.}
  \label{fig:polarizations}
\end{figure}

For example, for a GW propagating along the $z$-direction the GW in TT gauge will take the simple form
\begin{align}
 h_{ij} = A_{ij} \cos(k z) \,,
\end{align}
with
\begin{align}
 A_{ij} = \begin{pmatrix}
           h_+ & h_\times & 0 \\
           h_\times & - h_+ & 0 \\
           0 & 0 & 0
          \end{pmatrix} \,,
\end{align}
making the two polarization degrees $h_+$ and $h_\times$ explicit. Plugging this into Eq.~\eqref{eq:xi_eom} we see how test masses in the $x-y$ plane move in response to the GW. We obtain the names-giving $+$ and $\times$ pattern depicted in Fig.~\ref{fig:polarizations}.

Note that the appearance of $h^{TT}$ in the expression for the Newtonian force in the proper detector frame is due the convenient gauge invariance of the Riemann tensor.
The explicit form of the GW in the proper detector frame is more cumbersome to obtain, see Ref.~\cite{Rakhmanov:2014noa} for a derivation in the low frequency limit and Refs.~\cite{Berlin:2021txa,Domcke:2022rgu} for generalizations.


\section{Lecture II - Stochastic gravitational wave backgrounds}
\label{sec:lect2}

\subsection{The energy momentum tensor of GWs}
So far, we have considered GWs in flat (or FRW) space time. However, the background space time in our Universe is locally curved by the presence of astrophysical objects and dark matter halos. In this case, the previous procedure of identifying the homogeneous parts of the metric as background metric and the position dependent parts as GWs will no longer work. This raises the question of how to even define a GW in a general curved space time.

The key to answer this lies in a separation of scales. We will typically be interested in a situation where the background varies slowly (in time or space) compared to variations due to GW. Consider for a example a GW with wavelength $\lambda$ and background metric which varies on a typical scale $L \gg \lambda$. We can then choose a scale $d$ such that $\lambda \ll d \ll L$ to separate the GW contribution from the background: averaging over a domain of length $d$ gives the background part, subtracting this from the full expressions gives the GW part.

Let us perform this exercise on the left-hand side of Einstein's equation. Expanding in powers of the GW,
\begin{align}
 G_{\mu\nu} = \underbrace{G_{\mu\nu}^{(B)}}_{{\cal O}(h^0), L} + \underbrace{G_{\mu\nu}^{(1)}}_{{\cal O}(h^1), \lambda} + \underbrace{G_{\mu\nu}^{(2)}}_{{\cal O}(h^2), L \& \lambda} + ...
 \label{eq:Gexp2}
\end{align}
the first term has the characteristic length scale $L$, the second is dominated by the GW wavelength $\lambda$ (in Fourier space, $\vec k_{\lambda} + \vec k_L \simeq \vec k_\lambda$ since $k_\lambda \gg k_L$) and the third term contributes to both the short and the long wavelength since anti-aligned short wavelength modes can give a long wavelength contribution.

We will focus here on the long wavelength part of Eq.~\eqref{eq:Gexp2} which describes how the GW impacts the background metric. (The short wavelength part will describe the propagation of GWs in the curved background.) Given the expansion above, we can write Einstein's equation as
\begin{align}
 G_{\mu\nu}^B & = - \left[ G_{\mu\nu}^{(2)} \right]^L + 8 \pi G \left[ T_{\mu\nu} \right]^L \nonumber \\
 & = - \langle G_{\mu\nu}^{(2)} \rangle_d + 8 \pi G \langle T_{\mu\nu} \rangle_d \,,
\end{align}
where the angular brackets denote an average over a domain of size $d$. We see that at second order in $h$, the GWs impact the background metric through $\langle G_{\mu\nu}^{(2)} \rangle_d$. This suggest to interpret this term as the energy momentum tensor of GWs,
\begin{align}
 t_{\mu\nu} = - (8 \pi G)^{-1} \langle G_{\mu\nu}^{(2)} \rangle \,.
\end{align}
An explicit expression can be obtained by expanding Einstein's equation to second order in the GW amplitude. This is performed e.g.\ in Ref.~\cite{Maggiore:2007ulw} and after some rather lengthy algebra along the lines of Lecture~\ref{sec:lect1} one obtains
\begin{align}
  t_{\mu\nu} = (32 \pi G)^{-1} \langle \partial_\mu h_{\alpha \beta}^{TT} \partial_\nu h^{\alpha\beta}_{TT} \rangle \,.
\end{align}
In particular, this implies for the energy density of GWs,
\begin{align}
 \rho_\text{gw} = t_{00} =  (32 \pi G)^{-1} \langle  \dot h_{ij}^{TT} \dot h^{ij}_{TT} \rangle \,.
 \label{eq:energy}
\end{align}
Parametrically, $\rho_\text{gw} \sim h^2 \omega^2 M_{pl}^2$, with $M_{pl} = (8 \pi G)^{-1/2}$ the reduced Planck mass. As expected for a massless degree of freedom, this scales as radiation in an expanding universe, $\rho_\text{gw} \sim a^{-4}$.

\subsection{Stochastic GW backgrounds}
Stochastic gravitational wave backgrounds (SGWBs) arise from the superposition of GWs with different wave-vectors, frequencies and phases. They can be both of astrophysical or cosmological origin. Astrophysical SGWBs arise from the GWs emitted by unresolved sources, i.e.\ faint sources which are outside the detection volume for an individual detection but which nevertheless contribute to the signal observed in the detector. Cosmological SGWBs arise from events in our cosmic history that released significant amount of GWs (such as a first-order phase transition). To an observer today, they appear to be emitted from a spherical surface with a radius matching the cosmic time of the event, very similar to the cosmic microwave background. Arising from the superposition of many different GWs, SGWBs are typically isotropic, unpolarized and Gaussian, and they appear in the detector as an additional `noise' contribution.
Distinguishing this from instrumental noise is challenging. In ground-based interferometers such as LIGO, searches for SGWBs are done by cross-correlating different detectors utilizing that SGWBs give a correlated signal whereas the instrumental noise is expected to be (largely) uncorrelated across different detectors. Similarly, in pulsar timing arrays, time-delays from different pulsars are correlated to search for the characteristic imprint of GWs. We will discuss SGWBs and their imprints in pulsar timing arrays in more detail in Lecture~\ref{sec:lect4}.

Given what we have learned about propagation of GWs on sub- and super-horizon scales, it is useful to parametrize GWs as
\begin{align}
 h_{ij}^{TT}(\tau, \vec x) = \sum_{\lambda = +,\times} \int d^3 k \, h_\lambda(\vec k) {\cal T}_k(\tau) e_{ij}^\lambda(\hat k) e^{- i k x} + h.c. \,,
\end{align}
where $e_{ij}^\lambda$ is the polarization tensor, $e^\lambda_{ij}(\hat k) e^{ij}_{\lambda'}(\hat k) = 2 \delta_{\lambda \lambda'}$, $k_i e^{ij}_\lambda(\hat k) = 0$. Here we have factorized the Fourier coefficient into a primordial coefficient $h_\lambda(\hat k)$ parametrizing the GW spectrum at time $\tau_*$ and a transfer function ${\cal T}_k(\tau) = a(\tau_*)/a(\tau)$ describing the decay in the GW amplitude due to the expansion of the universe. Here $\tau_*$ is the time where the GW is sourced, or, for super-horizon GWs as sourced by cosmic inflation, the time when the GW enters the horizon.
Assuming homogeneity and isotropy,
\begin{align}
 \langle h_\lambda(\vec k) h_{\lambda'}(\vec k')\rangle = (2\pi)^3 \delta_{\lambda\lambda'} \delta(\vec k + \vec k') P_\lambda(|\vec k|) \,,
 \label{eq:isotropy}
\end{align}
we can then express the energy density of GWs, see Eq.~\eqref{eq:energy}, on sub-horizon scales as
\begin{align}
 \rho_\text{gw}(\tau_0) & = (32 \pi G \pi^2 a^2(\tau_0))^{-1} \int k^2 dk^3 \sum_\lambda P_\lambda(|\vec k|) \frac{a^2_*}{a^2_0}  \nonumber \\
 & = \rho_c \int d\ln k \, \frac{1}{\rho_c} \frac{\partial \rho_\text{gw}}{\partial \ln k} \equiv \rho_c \int d \ln k \, \Omega_\text{gw}(k)\,.
 \label{eq:Omegadef}
\end{align}
In the last line we have introduced the GW spectrum $\Omega_\text{gw}$, normalized to the critical density $\rho_c$. It can be read off by comparing to the integrand in the first line. Note that we have replaced the volume (or time) average in Eq.~\eqref{eq:energy} with the ensemble average~\eqref{eq:isotropy}, which implicitly assumes a very large amount of independent data realizations. For ground-based interferometers such as LIGO, this is completely justified: the detector is sensitive at frequencies around 100~Hz, and we have months of data in every data set. For pulsar timing arrays which target GWs with periods of up to $\sim$10~years, this is on the other hand not a very good approximation, resulting in a residual cosmic variance on the measurement, as we will discuss in Lecture~\ref{sec:lect4}.

\subsection{Characteristic frequencies of relic GWs}
Consider a GW observed today at frequency $f_0$. Taking into account redshift, this implies that the GW was sourced with a frequency $f_*$,
\begin{align}
 f_0 = f_* \frac{a_*}{a_0} \,, \quad f_* = (\epsilon_* H_*^{-1})^{-1}
\end{align}
with $\epsilon \lesssim 1$ parametrizing the GW source in units of the Hubble horizon $H_*$ at the sourcing time. Assuming for simplicity a radiation dominated universe, $H_*^2 = \pi^2 g_* T_*^4/(90 M_{pl}^4)$, we find
\begin{align}
 f_0 &\simeq  \epsilon_*^{-1} \left(\frac{T_*}{10^8~\text{GeV}} \right)~\text{Hz}  \label{eq:frel}\\
 t_* &\simeq 10^{-22} \, \epsilon_*^{-1} \left(\frac{\text{Hz}}{f_0} \right)^2 ~\text{s} \,.
\end{align}
Consequently, gravitational wave detectors sensitive at higher frequencies probe earlier times in cosmic history. In this sense, pulsar timing arrays, operating at nHz frequencies probe energy scales around the QCD scale. The upcoming laser interferometer space antenna LISA, most sensitive around mHz frequencies, will probe energy scales around the electroweak phase transition. Ground-based interferometers operating at around 100 Hz probe energy scales up to $10^8$~GeV. Of course, a crucial requirement for detecting the extremely weak signal of early universe gravitational waves is a sufficient detector sensitivity.

\subsection{Constraints on the amplitude of relic GWs}
Since gravitational waves contribute to the energy density of the universe as radiation, their total energy density is constrained by measurements of the expansion history of our universe. This is often parametrized in terms of an effective number of neutrinos $N_\text{eff}$. After electron decoupling, we can write the radiation energy density of the universe as
\begin{align}
 \rho_\text{rad} = \frac{\pi^2}{30}\left(2 + \frac{7}{4} N_\text{eff} \left(\frac{4}{11}\right)^{4/3}\right) T^4 \,.
\end{align}
The first term in the bracket is associated to the two degrees of freedom of the photon, the second term counts the effective number of neutrinos, with $N_\text{eff}^\text{SM} = 3.046$ in the SM. Measurements of the expansion history of our universe at BBN and CMB decoupling set bounds on any extra radiation, parametrized as $\Delta N_\text{eff}$~\cite{Planck:2018vyg,Pisanti:2020efz}. Applying this to GWs,
\begin{align}
 \rho_\text{gw} \leq \frac{7}{4}  \left(\frac{4}{11}\right)^{4/3} \Delta N_\text{eff} \, \rho_\gamma \,.
\end{align}
BBN and CMB constrain $\Delta N_\text{eff} \lesssim 0.2$, implying that the energy in gravitational waves can be at most about 10$\%$ of the photon energy density. Given that today, $\Omega_\gamma = \rho_\gamma/\rho_c \sim 10^{-5}$, this implies
\begin{align}
 \rho_\text{gw} \lesssim 0.1 \, \Omega_\gamma \, \rho_c \sim 10^{-6} \rho_c \,.
 \label{eq:BBN}
\end{align}
Assuming a SGWB with a broad spectrum (in units of logarithmic frequency), this is often phrased as a limit on the amplitude of the GW spectrum, $\Omega_\text{gw} \lesssim 10^{-6}$. Note that this limit applies only to GWs which were already present (and sub-horizon) at the time of BBN or CMB decoupling.

\section{Lecture III - Discovery opportunities with transient GWs}
\label{sec:lect3}

\begin{figure}
\centering
 \includegraphics[width = 0.5 \textwidth]{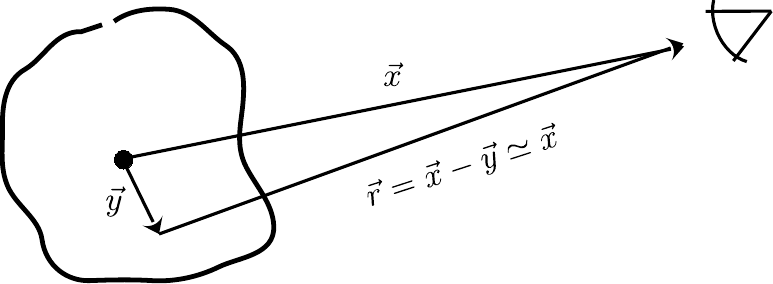}
 \caption{GW observed by an observer at location $\vec x$ far away from the source.}
 \label{fig:farfield}
\end{figure}

\subsection{Emission of GWs}
Consider an observer sufficiently far away from a GW source, as depicted in Fig.~\ref{fig:farfield}. Unless stated otherwise, we will in this lecture assume a flat background metric and neglect the expansion of the universe.  We can formally solve Eq.~\eqref{eq:wave_equation} using the retarded Green's function for the Klein Gordon equation, completely analogous to electromagnetism,
\begin{align}
 \bar h_{\mu\nu}(t, \vec x) = \frac{4 G}{r} \int d^3y \, T_{\mu\nu}(t - r, y)\,,
\end{align}
with $T_{\mu\nu}$ denoting the energy momentum tensor of the source. Far away from the source we will be able to use TT gauge, so it suffices to focus on the spatial components when evaluating this expression. We can make progress by using the following relation, which we prove below:
\begin{align}
 \int d^3 y \, T^{ij}(y) = \frac{1}{2} \partial_0^2 \int d^3 y \, y^i y^j T_{00}(y) \,.
 \label{eq:Trel}
\end{align}

\begin{quote}
\small{\noindent \textit{Proof:} Using energy-momentum conservation, $\partial_\mu T^{\mu\nu} = 0$,
\begin{align}
 \partial_\mu T^{0\mu} = 0 = \partial_0 T^{00} + \partial_k T^{0k} \,. \nonumber
\end{align}
Then acting with a time derivative we obtain
\begin{align}
 \partial_0 T^{00} = - \partial_k \partial_0 T^{0k} = \partial_k \partial_l T^{kl}\,, \nonumber
\end{align}
where in the second step we have used energy-momentum conservation again. Now we multiply with $y^i y^j$,
\begin{align}
 y_i y_j \partial_0^2 T^{00} = y_i y_k \partial_k \partial_l T^{kl} \rightarrow 2 \partial T^{ij}\,, \nonumber
\end{align}
where the arrow in the last step indicates an integration by parts under the $d^3y$ integral. This proves the relation~\eqref{eq:Trel}.}
\end{quote}

\noindent With this result, we can express
\begin{align}
 \bar h_{ij}(t, \vec x) = \frac{2 G}{r} \partial_0^2 \int d^3y \; y_i \, y_j \, T_{00}(t-r,y) \,,
\end{align}
where we recognize the integral as the tensor moment of the source. It will sometimes be convenient to instead work with the trace-free version of this quantity, the quadrupole moment,
\begin{align}
{I}_{kl} = \int d^3y \, (y_k \, y_l - \frac{1}{3} y^2 \delta_{kl}) \, T_{00} \,.
\end{align}
In TT gauge, we can then write,
\begin{align}
 h_{ij}^{TT}(t, \vec x)= \frac{2G}{r} \Lambda_{ij,kl} {\ddot I}_{kl}(t-r) \,,
 \label{eq:sourcing}
\end{align}
where $\Lambda_{ij,kl}$ is a projector which retains only the transverse traceless components.\footnote{
Such a projector can be explicitly constructed as $ \Lambda_{ij,kl} = P_{ik} P_{jl} - P_{ij} P_{kl}/2$ where $P_{ij} = \delta_{ij} - n_i n_j$ is the transverse projector with respect to the unit vector of the GW propagation direction, $\vec n = \vec k/k$.}
Equation~\eqref{eq:sourcing} demonstrates that GWs are sourced by objects which feature accelerated quadrupole moments.

\subsection{Einstein's quadrupole formula}
We now turn to the power emitted by a GW source. Consider a sphere with volume $V$ containing a GW source, with a radius much larger than the source dimensions. The energy of GWs in this volume is $\epsilon_\text{gw} = \int_v d^3x \, t^{00}$ with $t_{\mu\nu}$ the energy momentum tensor of GWs. Using energy-momentum conservation, $\partial_\mu t^{\mu 0} = 0$, we can express the emitted power as
\begin{align}
 P_\text{gw} = \frac{d \epsilon_\text{gw}}{dt} = - \int_V d^3 x \, \partial_i t^{0i} = - \int_S d^2S \,  n_i t^{0i} = - r^2 \int_S d\Omega \, t^{0r} \simeq r^2 \int_S d\Omega \, t^{00} \,,
\end{align}
where $\hat n$ is a normal vector to the surface $S$ enclosing the volume $V$. In the last step, we have made use of
$ t^{0r} \propto \langle \partial_0 h_{ij}^{TT} \, \partial_r h_{ij}^{TT} \rangle $,
with $r \,  h_{ij}^{TT}$ a function that depends on space-time only through $t - r/c$. Taking partial derivatives yields $\partial_r h_{ij}^{TT} = - \partial_0 h_{ij}^{TT} + {\cal O}(1/r^2)$.

Inserting the expression~\eqref{eq:energy} for the GW energy density and the solution~\eqref{eq:sourcing} for the GW in the far-field regime we obtain
\begin{align}
 P_\text{gw} &= \frac{r^2}{32 \pi G} \int d\Omega \langle \dot h_{ij}^{TT} \dot h_{ij}^{TT} \rangle \nonumber  \\
 & = \frac{G}{8 \pi} \int d\Omega \, \Lambda_{ij,kl}(\hat n) \, \langle \dddot I_{ij} \dddot I_{kl} \rangle \nonumber \\
 & = \frac{G}{5} \langle \dddot I_{ij} \dddot I_{kl} \rangle \,.
 \label{eq:Equadrupole}
\end{align}
In the last step we have performed the angular integral $\int d\Omega \, \Lambda_{ij,kl}(\hat n) = \frac{2\pi}{15} (11 \delta_{ik} \delta_{lj} - 4 \delta_{ij} \delta_{kl} + \delta_{il} \delta{jk})$. This expression for the total power of the emitted GWs is referred to as Einstein's quadrupole formula.

\subsection{Binary systems}
Let us consider a bound system of two point-like objects with masses $m_{1,2}$ located at positions $\vec r_{1,2}$. Introducing the reduced mass $\mu = m_1 m_2/(m_1 + m_2)$, the total mass $M = m_1 + m_2$ and the distance between the two objects $\vec r = \vec r_2 - \vec r_1$,  we can express this as an effective one-body problem where an object of reduced mass $\mu$ is rotating around the center of mass with a frequency given by Keppler's law, $\omega_\mu^2 = G M /r^3$. This system models binary systems of astrophysical objects, such as black holes or neutron stars, as long as the separation between the objects is large enough such that the internal structure of the objects is irrelevant. For simplicity, we will take the orbit to be circular.

In order to compute the GW emission from this system using Eq.~\eqref{eq:sourcing}, we need to compute the quadrupole moment. Taking the binary to be in the $x-y$-plane,
\begin{align}
 x(t) = r \cos (\omega_\mu t) \,, \quad y(t) = r \sin(\omega_\mu t) \,, \quad I_{ij} = \mu (r_i r_j - \frac{1}{3} r^2 \delta_{ij})
\end{align}
we find for an observer along the $z$-axis at a distance $d$,
\begin{align}
 h_+^{TT}(t, d \hat e_z) & = \frac{G}{d}(\ddot I_{xx} - \ddot I_{yy})  \nonumber \\
 & = - \frac{4}{d} (G M_c)^{5/3} \omega_\mu^{2/3} \cos(2 \omega_\mu t) \,,
 \label{eq:hinsp}
\end{align}
where we have introduced the so-called chirp mass, $M_c = (\mu^3 M^2)^{1/5}$. Note that the GW frequency is twice the orbital frequency of the binary system, $2 \pi f_\text{gw} = \omega_\text{gw} = 2 \omega_\mu$.
For cosmological distances, we will need to replace the distance $d$ with the luminosity distance $d_L = (1 + z) d$, and take into account the frequency redshift. In general, we will also have to account for the inclination angle of the observer with respect to the plane of the galaxy as well as the ellipticity of the orbit, see e.g.~Ref.~\cite{Maggiore:2007ulw} for expressions containing these correction factors. Finally note that this expression was obtained in the Newtonian limit. Close to the merger, relativistic corrections and strong gravitational effects become important, which can be treated perturbatively by expansions in powers of the velocity (post-Newtonial expansion) and the gravitational constant (post-Minkowskian expansion).

\begin{figure}
 \centering
 \includegraphics[width = 0.7\textwidth]{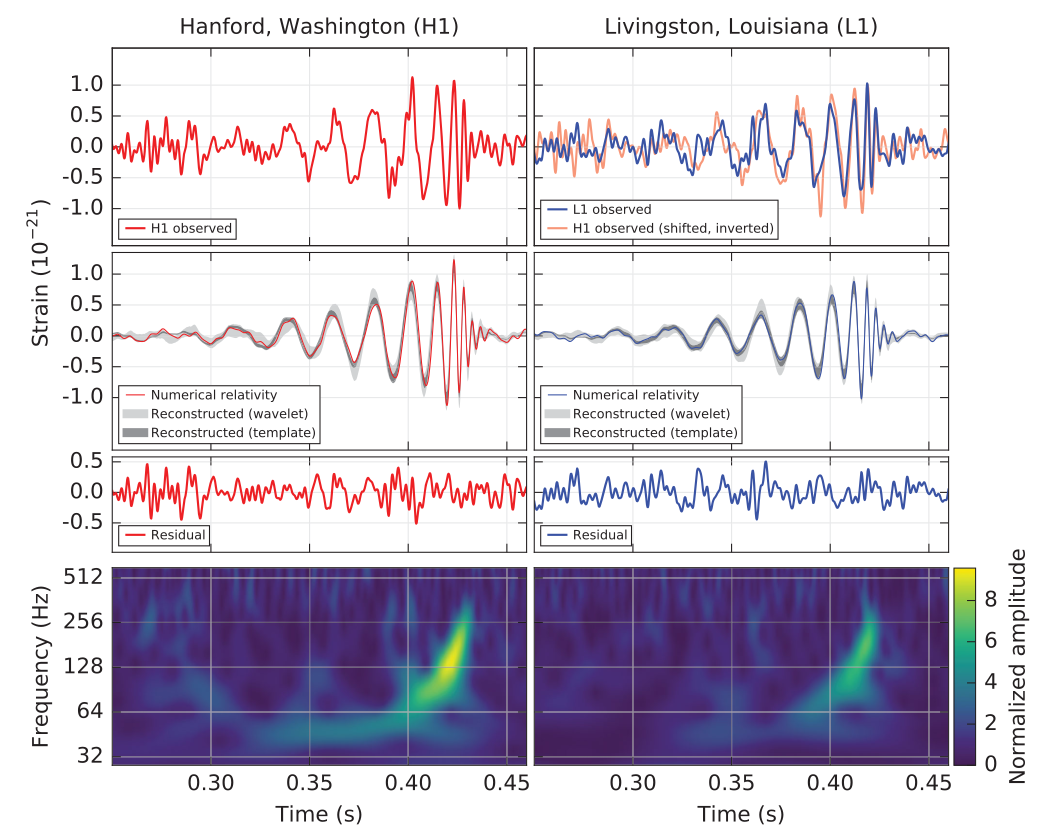}
 \caption{First direct gravitational wave detection by the LIGO/Virgo collaboration (GW150914), taken from~\cite{LIGOScientific:2016aoc}.}
 \label{fig:LIGOdetection}
\end{figure}

Neglecting the emission of GWs, the orbit of these objects is stationary. To self-consistently account for the energy loss due to the emission of GWs, we compute the power emitted into GWs by inserting Eq.~\eqref{eq:hinsp} into Einstein's quadrupole formula~\eqref{eq:Equadrupole},
\begin{align}
 P_\text{gw} & = \frac{G}{5} \langle \dddot I_{xx}^2 +  \dddot I_{yy}^2 + 2 \dddot I_{xy}^2 \rangle \nonumber \\
 & = \frac{2 G}{5} (4 \mu r^2 \omega_\mu^3)^2 \langle \cos^2(2 \omega_\mu t) + \sin^2(2 \omega_\mu t) \rangle \nonumber \\
 & = \frac{32 \, G^{7/3}}{5} {\cal M}_c^{10/3} \omega_\mu^{10/3} \,.
\end{align}
This energy loss leads to a slowly shrinking orbit for the binary. With the total energy of a virialized orbital system given by $E_\text{orbit} = - G m_1 m_2/(2 r) = - \frac{1}{2} G^{2/3} \omega_\mu^{2/3} {\cal M}_c^{5/3}$ we obtain
\begin{align}
 \frac{dE_\text{orbit}}{dt} & = - \frac{1}{3}  G^{2/3} \omega_\mu^{-1/3} {\cal M}_c^{5/3} \dot \omega_\mu \nonumber \\
 & = - P_\text{gw} = - \frac{32 \, G^{7/3}}{5} {\cal M}_c^{10/3} \omega_\mu^{10/3}\,.
\end{align}
The energy loss due to GWs thus implies a shrinking orbit and correspondingly and increasing orbital and GW frequency,
\begin{align}
 \dot f_\text{gw} = \frac{96}{5} \pi^{8/3} \left( G M_c \right)^{5/3} f_\text{gw}^{11/3} \,.
\end{align}
According to Eq.~\eqref{eq:hinsp}, an increasing frequency implies an increasing GW amplitude, $h \propto f_\text{gw}^{2/3}$. This explains the characteristic 'chirp' signals of merging compact objects, see Fig.~\ref{fig:LIGOdetection}.

Observing such a signal over some period of time gives us access to three observables: amplitude $h$, frequency $f_\text{gw}$ and the change in frequency $\dot f_\text{gw}$. This allows us infer three parameters, typically the chirp mass $M_c$, the size of the binary $r$ and the luminosity distance $d_L$. There are however many other relevant parameters: the mass ratio $m_1/m_2$, the sky position, the inclination angle under which the observer sees the binary $\theta$, the ellipticity of the orbit, the spins of the two objects $s_{1,2}$ and the deformability of the objects $\Lambda_{1,2}$. On top of these 15 parameters, additional parameters may be added to describe possible new physics playing a role in this system. To break these degeneracies, we need access to more observables which we can obtain by studying the merger phase (going beyond the Newtonian limit presented here), the post-merger ring-down phase, and by using multiple detectors to access polarization information.

\subsection{Ground-based interferometers: LIGO/Virgo/Kagra (LVK)}


Interferometers are sensitive to GWs through the relative change in proper distance in the two arms, which changes the interference pattern observed after interfering the photons arriving back at the beam splitter from the two arms. In TT gauge,
\begin{align}
 ds^2 & = - dt^2 + (\delta_{ij} + h_{ij})dx^i dx^j \nonumber \\
 & = - dt^2 + (1 + h_{ij} \hat l^i \hat l^j) d\lambda^2 \nonumber \\
 & = - dt^2 + (1 + \frac{1}{2} h_{ij} \hat l^i \hat l^j)^2 d\lambda^2 \,,
\end{align}
with $\hat l$ the unit vector along the interferometer arm with length $L$ and $\lambda = [0,L]$ parametrizing the distance along this arm. From this we read off the time delay due to the GW as
\begin{align}
 \Delta T(t) & = \frac{1}{2} \hat l^i \hat l^j \int_0^L h_{ij}(t + \lambda, \lambda \, \hat l) d\lambda\\
 & \equiv L \int d^3 k \sum_\lambda F_\lambda(\vec k)\, h_\lambda(\vec k) \,,
 \label{eq:time_delay}
\end{align}
where we have introduced the sinlge-link detector response function $F_\lambda(\vec k)$. To get the full detector response function, we add the analogous expression for the return trip of the photon and subtract the result obtained from the second interferometer arm.
This time delay translates into a phase difference between the two lasers beams arriving back at the beam splitter, which is read out in an interferometric measurement. Due to the hierarchy between the arm length (4~km for LIGO, multiplied by a factor of about $10^3$ to account for the finesse of the Fabry-Perot cavity) and the optical wavelength, this measurement is sensitive to changes in the mirror positions less than the size of a proton!

To date the LVK collaboration has observed about a hundred black hole (BH) mergers, and a few neutron star (NS) and BH-NS mergers. Below, we will look at some implications for fundamental physics obtained from this data. The next generation of ground-based telescopes, such as the Einstein Telescope~\cite{Maggiore:2019uih} and the Cosmic Explorer~\cite{Reitze:2019iox}, will significantly increase this number, pushing the red-shift reach for BH mergers beyond star formation time: this implies a sensitivity to be able to see all stellar-origin black hole mergers in this mass range in the observable universe.

\paragraph{Test of general relativity with neutron star mergers (GW170817).}
While merging BHs are not expected to emit any significant electromagnetic (EM) signal, merging NSs emit EM radiation in form of a directed beam. If we happen to be lucky enough to find ourselves within the opening angle of this beam, we can observe an optical counterpart to the GW signal of merging NSs. Remarkably, this was the case for the event GW170817~\cite{LIGOScientific:2017vwq,LIGOScientific:2017ync}.

The GW signal was observed by Ligo and Virgo and lasted about 100~s. The resulting sky localization indicated a fairly broad patch of the sky as probable direction of the signal. A follow-up campaign by dozens of EM telescopes was able to detect and follow the EM signal from about a second after the GW signal up days later. The probable host galaxy was identified as NGC4993, enabling a determination of its red shift. Comparing the arrival time of the GW signal and the arrival time of the EM signal resulted in a stringent bound on the velocity of GWs,
\begin{equation}
 - 3 \cdot 10^{-15} \leq \frac{v_\text{gw} - c}{c} \leq 7 \cdot 10^{-16} \,,
\end{equation}
consistent with the prediction of general relativity (GR) of $v_\text{gw} = c$, and resulting in stringent constraints on e.g.\ models of massive gravity~\cite{Creminelli:2017sry,Baker:2017hug,Ezquiaga:2017ekz}.

\begin{figure}
 \centering
 \includegraphics[width = 0.7\textwidth]{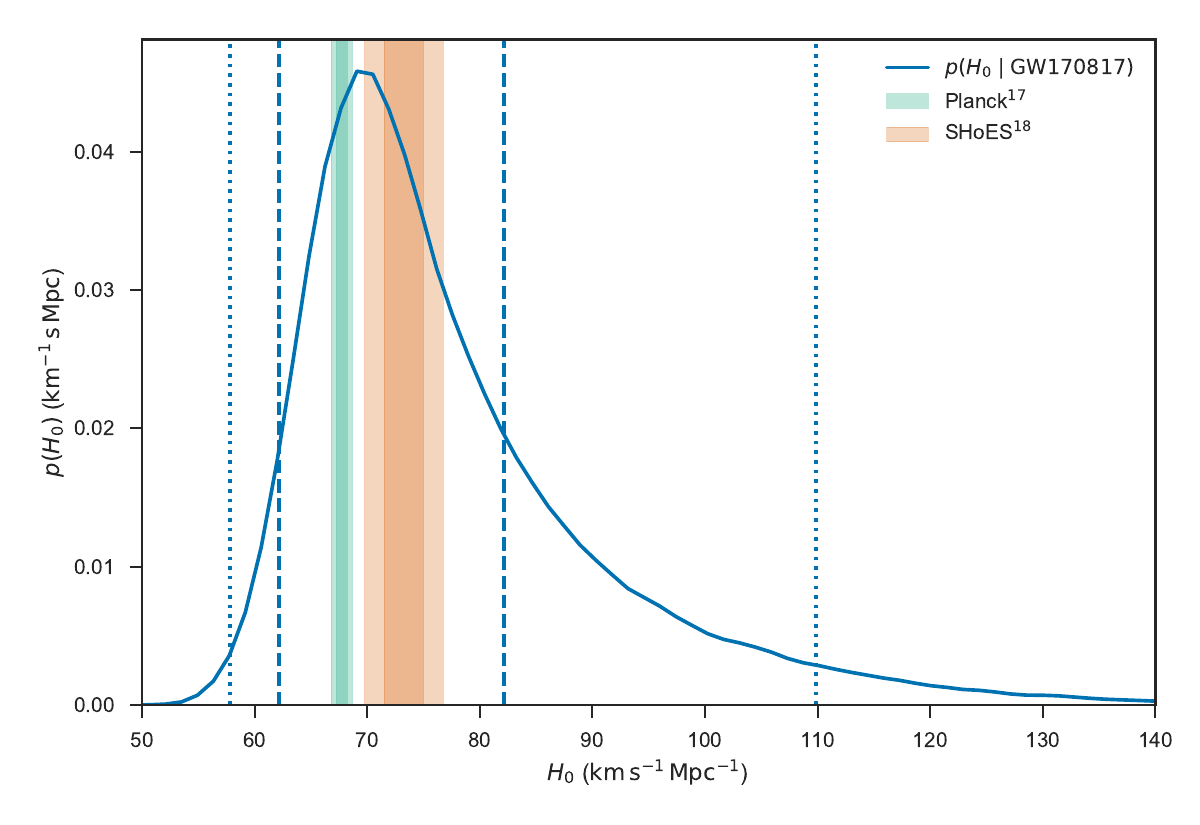}
 \caption{First measurement of the Hubble parameter with GWs, figure taken from \cite{LIGOScientific:2017adf}.}
 \label{fig:hubble}
\end{figure}

\paragraph{Measuring the Hubble parameter with neutron star mergers  (GW170817).}
The same event also allowed for a first measurement of the Hubble parameter with GWs~\cite{LIGOScientific:2017adf}. From the GW signal, one can infer the luminosity distance $d_L = (1 + z) d$. The identification of the host galaxy allowed to determine the redshift $z$, making this event a `standard siren' (analogous to the standard candles used to create cosmic distance ladders). Using Hubble's law,
\begin{align}
 z = H_0 \, d_L \,,
\end{align}
this give a measurement of the Hubble parameter $H_0$. The result is shown in Fig.~\ref{fig:hubble}. While this single measurement does not provide enough accuracy to resolve the long standing tension between the value infrerred from the CMB (Planck) and through distance ladder measurements (such as SHoES), this will change if we observe more such events. An alternative approach is the dark standard siren approach: For a sufficiently large sample of mergers, the posteriors for the sky localization can be correlated with galaxy catalogs to assign probably host galaxies~\cite{LIGOScientific:2021aug}. This allows a statistical determination of the Hubble parameter even without the observation of optical counterparts.

\begin{figure}
 \centering
 \includegraphics[clip, trim=0cm 0cm 18cm 0cm,width = 0.7\textwidth]{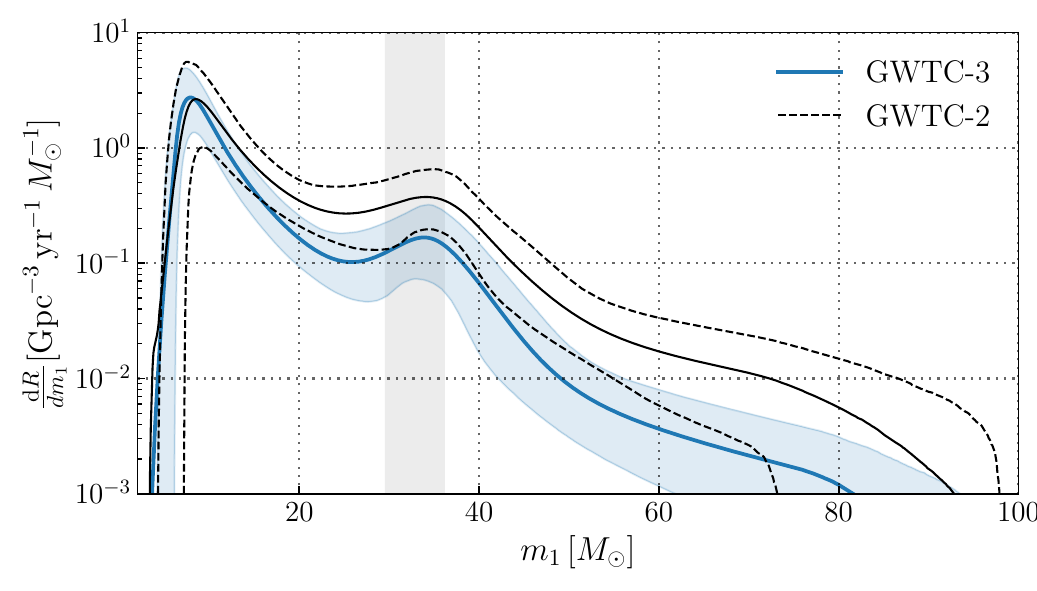}
 \caption{Inferred black hole mass distribution from observing runs 2 and 3 of the LVK collaboration, figure taken from \cite{KAGRA:2021duu}.}
 \label{fig:masses}
\end{figure}

\paragraph{Test of star formation (BH formation).}
Fig.~\ref{fig:masses} shows the BH mass distribution inferred from LVKs second and third observing run~\cite{KAGRA:2021duu}. The observed distribution shows a remarkable amount of substructure and carries information about the history of star formation. We do not expect to form stellar origin black holes below the Chandrasekhar limit of $1.4~M_\odot$. A detection of a BH below this limit (if solidly distinguishable from a NS) would be a smoking gun of primordial black hole production, with exciting consequences for cosmology. However, beyond this there is a noticeable lack of BH observations up to about $6~M_\odot$. This may be tied to an underestimated selection bias or to the dynamics of core-collapse supernovae. Similarly, the peak at around $35~M_\odot$ might be associated to stellar physics, astrophysical environments, or supernova dynamics. At the high mass end, beyond about $70~M_\odot$, there is instead an excess of BHs compared to expectations from late-time stellar evolution. In this mass range, we expect pair-instability supernovae, which do not result in a BH remnant. It has been proposed that these BHs are generated by accretion onto or multiple mergers of lighter BHs. In summary, observations so far are pointing towards the existence of several different relevant BH formation channels, with the wealth of new data promising an improvement of our understanding of star formation and already giving rise to a range of new questions.

\subsection{Interferometers in space: LISA}
The Laser Interferometer Space Antenna (LISA)~\cite{LISA:2017pwj,Colpi:2024xhw} is a space-based mission scheduled for launch in the 2030s. Three satellites span an equilateral triangle with 2.5 million kilometer arm length, each of them on an orbit around the sun, trailing the earth. They will constitute an interferometer with a peak sensitivity in the mHz range. Beyond a rich astrophysical program, key science goals of LISA include important probes of fundamental physics: the measurement of the Hubble parameter with supermassive BH binaries (standard sirens) at high redshift, probing the near field region of BHs through extreme mass ratio inspirals (EMRIs), bound systems in which a stellar origin black hole acts as a test mass to map out the metric of a supermassive black hole, and finally the search for stochastic gravitational wave backgrounds from the early universe, see lecture~\ref{sec:lect4}.

\section{Lecture IV - Discovery opportunities with stochastic gravitational wave backgrounds}
\label{sec:lect4}

\subsection{Pulsar timing arrays}
In 2023, pulsar timing arrays (PTAs) across the world announced evidence of a first observation of GW signal that can be interpreted as a SGWB~\cite{NANOGrav:2023gor,EPTA:2023fyk}. Pulsar timing arrays observe dozens of  galactic millisecond pulsars. These pulsars emit a beam of EM radiation sweeping through the Universe as the pulsars rotate. They are at cosmic light houses, seen as regular, pulsed radio signals. A passing gravitational wave will modify the arrival time of these pulses compared to prediction of the pulsar model, which takes into account physical processes within the pulsar, as well as the relative motion of pulsar and the observatory. The time delay $\Delta T$ is obtained completely analogously to our computation for interferometers and is given by Eq.~\eqref{eq:time_delay}. The pulsar timing array literature commonly uses redshifts,\footnote{PTAs measure frequencies as low as $1/T_\text{obs}$ with $T_\text{obs}$ the observation time. This implies that in the low frequency regime, time-averaging the data does not (completely) remove the GW imprint. For a plane GW, this results in a dependence of the measured time delays on the unknown GW phase, or in other words, the timing model must account for this unknown phase parameter. This is circumvented by working with redshifts.} obtained as
\begin{align}
 z(t) = \frac{d}{dt} \Delta T(t) \,.
\end{align}
For a pulsar with label $\alpha$ with the unit vector $\hat p$ denoting its sky-localization we obtain
\begin{align}
 z_\alpha(t) = \int d^3k \sum_\lambda F_\lambda^\alpha(\hat k) h_\lambda(t, \vec 0) \quad  \text{with} \quad  F_\lambda^\alpha(\hat k) \simeq  \frac{\hat p^i \hat p^j}{2(1 + 2 \pi \hat k \cdot \hat p)} e_{ij}^\lambda(\hat k) \,.
 \label{eq:pulsar_response}
\end{align}
As above, $\vec k$ denotes the GW wave vector.
In evaluating $F_\lambda^\alpha(\hat k)$ we have used that in the frequency range relevant for pulsar timing arrays (set by the inverse of the observation time, $\sim 10$~years), the GW wavelength is much shorter than the distance to the pulsars:
\begin{align}
 10 \text{ years} \sim 1/(3 \text{ nHz}) \sim 3 \text{ pc} \ll \text{ kpc} \,,
 \label{eq:pta_scales}
\end{align}
where kpc is the relevant scale for the galactic distances to the pulsars. Consequently, in the integration in Eq.~\eqref{eq:time_delay}, the contribution from the upper integration boundary (``pulsar term'') is suppressed, and we have dropped it in the expression above.

\begin{figure}
\centering
 \includegraphics[width = 0.6 \textwidth]{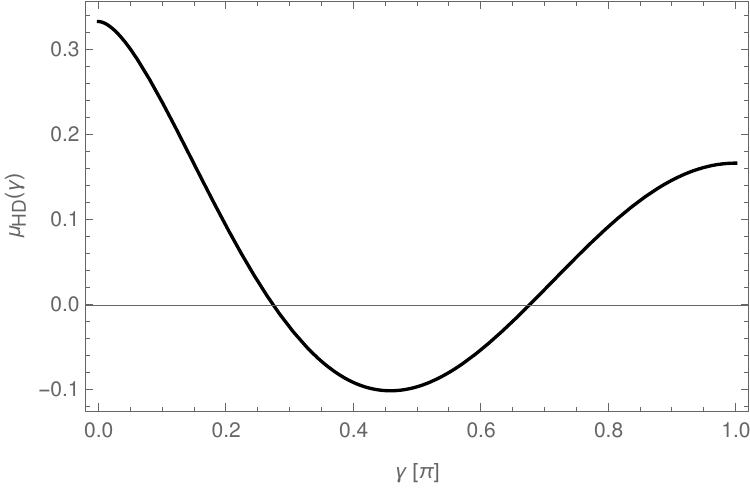}
 \caption{Cross-correlation of time delays from  pulsars separated by an angle $\gamma$ (Hellings-Down curve), taking the ensemble average over an unpolarized and isotropic SGWB.}
 \label{fig:HD}
\end{figure}

The observable of interest is the cross-correlation of the time delays from two pulsars $\alpha$ and $\beta$,
\begin{align}
 s_{\alpha \beta} = \langle z_\alpha(t) z_\beta(t) \rangle \,.
\end{align}
For an unpolarized, isotropic SGWB we can use Eq.~\eqref{eq:isotropy} to evaluate this explicitly,\footnote{In doing so, we will replace the volume average of Eq.~\eqref{eq:energy} with the ensemble average appearing in \eqref{eq:isotropy}. Given Eq.~\eqref{eq:pta_scales}, these are not equivalent. The result obtained in Eq.~\eqref{eq:HD} is in fact the mean for the Hellings-Down curve obtained by averaging over many (hypothetical) universes, given a single realization of the Universe this measurement is instead subject to cosmic variance~\cite{Allen:2022dzg}.
}
\begin{align}
 s_{\alpha \beta} & = \int k^2 dk \, P_\lambda(|\vec k|) \int d\hat k \sum_\lambda F_\alpha^\lambda(\hat k) F_\beta^\lambda(\hat k)  \label{eq:pta_corr}\\
 & \equiv \int k^2 dk \, P_\lambda(|\vec k|) \, (4 \pi \mu_\text{HD}(\gamma)) \,.
\end{align}
In the second line we have introduced $\mu_\text{HD}$,  the Hellings-Down (HD) correlation as a function of the angle between the two pulsars, $\cos \gamma = \hat p_\alpha \cdot \hat p_\beta$. Inserting Eq.~\eqref{eq:pulsar_response}, we obtain
\begin{align}
 \mu_\text{HD}(\gamma) = \frac{1}{4} + \frac{1}{12} \cos \gamma + \frac{1}{2}(1 - \cos \gamma) \ln\left( \frac{1 - \cos\gamma}{2} \right) \,, \label{eq:HD}
\end{align}
displayed in Fig.~\ref{fig:HD}. As an aside, note that we obtain the same result by assuming a gravitational plane wave and assuming a large amount of pulsars distributed uniformly across the sky: in this case the angular integral in Eq.~\eqref{eq:pta_corr} is performed over the pulsar locations instead of the GW direction.
For more a more detailed discussion on detector response functions for SGWB searches see Ref.~\cite{Romano:2016dpx}.

Significant evidence for a HD correlation was reported by different pulsar timing array collaborations in 2023, pointing towards a first discovery of a SGWB. If confirmed, the most likely explanation are GWs from supermassive black hole mergers, i.e.\ an extragalactic signal due to the merger of supermassive black holes in the course of galaxy mergers. As pulsar timing arrays collect more data, we will learn more about the frequency distribution of this signal (SGWB or individual source, more precise information on the spectral shape of a SGWB component) and about its sky distribution~\cite{Mingarelli:2013dsa}. This will enable us to conclusively determine if this is indeed a SGWB and if it is of astrophysical (relatively anisotropic) or cosmological (typically very isotropic) origin~\cite{Depta:2024ykq}.

\subsection{Cosmological sources}
With a possible first glimpse of a SGWB observed at timing arrays, and the expectation to see (at least astrophysical) SGWBs both at LISA and in the next generation of ground-based interferometers, let us turn to cosmological sources which could contribute SGWBs. We will distinguish transient sources, which are cosmic events lasting less than a Hubble time, and continuous sources, which last much longer than a Hubble time. The former lead to peaked spectra with a peak frequency given by Eq.~\eqref{eq:frel}, the latter lead to much broader spectra, which can extend over several orders of magnitude in frequency. For a more detailed review of cosmological sources, see e.g.~\cite{Caprini:2018mtu}.

\subsubsection{Transient sources}

\paragraph{Cosmic gravitational microwave background (CGMB).}
The hot primordial plasma of the early universe emits EM radiation, which decouples at a temperature of around eV, when electrons and protons combine into neutral hydrogen. This radiation is known as the cosmic microwave background, described by a black body spectrum peaked at radio frequencies today. Similarly, the hot primordial plasma can emit gravitational radiation. The decoupling temperature of such radiation from the thermal plasma is the Planck scale. The maximal temperature ever reached by a thermal plasma in the post-inflationary universe is given by the reheating temperature after inflation.\footnote{The GW production is thus heavily UV dominated, which is why we categorize the CGMB as a transient source.
Note that larger temperatures can be reached during the reheating process by thermal plasmas which constitute only a small fraction of the total energy density of the Universe at that time. GW emission from these is suppressed by this small fraction as well as by subsequent the redshift during the reheating process.} Given the bound on the energy scale of inflation coming from the non-observation of gravitational waves in the CMB, this temperature is bounded from above, $T_\text{rh} \lesssim  10^{-2} \, M_\text{Pl} \simeq 10^{16}$~GeV. This leads to a suppression of the GW spectrum compared to the CMB~\cite{Ghiglieri:2015nfa},
\begin{align}
 \Omega_\text{gw}^\text{CGMB}(f) \sim \frac{T_\text{rh}}{M_{pl}} \, \Omega^\text{CMB}_\gamma(f) \,.
\end{align}
Just as the CMB, this spectrum is peaked at radio frequencies. Combined with the upper bound on its amplitude, this signal is unfortunately out of reach of any currently envisaged GW detectors.

\paragraph{First order phase transitions.}
The Standard Model (SM) features two phase transitions that occurred as the hot primordial plasma cooled in the expanding Universe: QCD confinement and electroweak symmetry breaking. Both of these phase transitions are cross-over phase transitions, and do not source any significant amount of GWs. However, extensions of the SM may feature more violent, first order phase transitions. This can be achieved at the electroweak scale by considering e.g.\ extended Higgs sectors, at the QCD scale if large chemical potentials were present, or one may consider additional symmetry breaking processes at higher temperatures or in a dark sector. In a first order phase transition, bubbles of true vacuum nucleate, expand and collide. This leads to sound waves and possibly turbulence in the primordial plasma. These processes are highly anisotropic and involve a significant fraction of the energy density of the universe, ideal conditions for strong gravitational wave signals. Much work has gone into understanding these processes in detail and into predicting the resulting GW signal, see e.g.~\cite{Caprini:2015zlo,Caprini:2019egz} for a review. Here, we restrict ourselves to a simple parametric estimate to get a rough idea of the expected magnitude of these signals.

Given the Lagrangian of a given BSM model, one can compute the bubble nucleation rate $\Gamma$  and the potential difference between the false and the true vacuum, $\rho_\text{vac}$. From this, one obtains the effective parameters,
\begin{align}
 \beta = \frac{\dot \Gamma}{\Gamma} \qquad \text{and} \qquad \alpha = \frac{\rho_\text{vac}}{\rho_\text{rad}} \,,
\end{align}
where $1/\beta$ is roughly the duration of the phase transition and $\alpha$ is the energy release of the phase transitions in units of the average energy density of the plasma $\rho_\text{rad}$.

Using Eq.~\eqref{eq:wave_equation}, we estimate
\begin{align}
 \beta^2 h_{ij} \sim \beta \, \dot h_{ij} \sim 16 \pi G \, T_{ij}^\text{(an)} \,,
\end{align}
which gives for the GW energy density~\eqref{eq:energy} at the time of the phase transition,
\begin{align}
 \rho_\text{gw}(\tau_*) \sim \frac{\dot h_{ij}^2}{32 \pi G} \sim \frac{8 \pi G \, (T_{ij}^\text{(an)})^2}{\beta^2} \,.
\end{align}
Expressing
\begin{align}
 |T_{ij}^\text{(an)}| = \frac{|T_{ij}^\text{(an)}|}{\rho_\text{vac}} \,  \frac{\rho_\text{vac}}{\rho_\text{rad}} \,  \rho_\text{rad} \equiv \kappa \, \alpha \, \rho_\text{rad} \,,
\end{align}
and using the Friedmann equation $3 H_*^2/(8 \pi G) = \rho_\text{rad}$,
we obtain for the GW energy density today,
\begin{align}
 \rho_\text{gw} \sim \rho_\gamma \left( \frac{H_*}{\beta} \right)^2 \, (\kappa \alpha)^2 \,.
\end{align}
The latter two factors are constraint to be less than unity, but for a sufficiently strong and slow phase transition they can come close to unity. This suggests that first order phase transitions can generate GW signals that saturate the BBN bound~\eqref{eq:BBN}.
This gives a rough estimate of the peak amplitude, $\Omega_\text{gw}(f_p) \sim \rho_\text{gw}/\rho_c$, with the peak frequency $f_p$ given by Eq.~\eqref{eq:frel}.

\subsubsection{Continuous sources}

\paragraph{Inflation.}
The exponential growth of the universe during cosmic inflation predicts that quantum vacuum fluctuations get stretched to superhorizon scales. On superhorizon scales these fluctuations freeze (see Lecture~\ref{sec:lect1}), until they re-enter the horizon during radiation or matter domination. The scalar fluctuations re-entering the horizon around CMB decoupling lead to the temperature anisotropies in the cosmic microwave background (CMB). Fluctuations entering earlier (later) induce density perturbations at smaller (larger) scales. Tensor perturbations (i.e.\ primordial gravitational waves) have not been observed yet, with the most stringent constraints coming from the search for polarization (B-modes) in the CMB. In single-field slow-roll inflation, the GW power spectrum is obtained as
\begin{align}
 P_\lambda(|\vec k|)= \left(\frac{2}{M_P}\right)^2 \frac{H^2_\text{inf}}{2 \, k^3}\,,
\end{align}
with $H_\text{inf}$ denoting the (approximately constant) Hubble parameter during inflation. This result is often quoted in terms of the scale invariant power spectrum
\begin{align}
 \Delta_t^2 = \frac{k^3}{2 \pi^2}\sum_\lambda P_\lambda(|\vec k|)  = r \, \Delta_s^2\,,
\end{align}
with $\Delta_s^2 \sim 2 \cdot 10^{-9}$ the amplitude of the scaler power spectrum observed in the CMB temperature fluctuations and $r \lesssim 0.06$ the tensor-to-scalar ratio~\cite{Planck:2018vyg}.
Plugging this into Eq.~\eqref{eq:Omegadef} with $a_* = a_k(\tau_*) = k/H(\tau_*)$ denoting the scale factor when the mode $k$ re-enters the horizon yields
\begin{align}
 \Omega_\text{gw}(k) = \frac{\Delta_t^2}{12} \left(\frac{k}{a_0 H_0}\right)^2 \left(\frac{a_*}{a_0}\right)^2
 = \frac{\Delta_t^2}{12} \left(\frac{k}{a_* H_*}\right)^2 \frac{a_*^4 H_*^2}{a_0^4 H_0^2} \,.
 \label{eq:inflation}
\end{align}
The first factor is approximately constant, bounded by above from CMB observations. The second factor gives $1$ by definition of $a_*$. The remaining task is to evaluate the last factor, which contains the information on the time of horizon re-entry and the subsequent cosmological history. In radiation domination, this is particular simple, since $H^2 \propto \rho_\text{rad} \propto a^{-4}$, and hence
\begin{align}
 \frac{a_*^4 H_*^2}{a_0^4 H_0^2} \sim \frac{a_\text{eq}^4 H_\text{eq}^2}{a_0^4 H_0^2} \sim \Omega_\gamma\,,
\end{align}
where the subscript `eq' indicates matter-radiation equality and we have neglected any changes in the number of degrees of freedom in the thermal bath. We conclude that for modes re-entering the horizon during radiation domination the GW spectrum is scale-invariant, $\Omega_\text{gw} \sim \Delta_t^2 \Omega_\gamma/12 \lesssim 10^{-16}$. In matter domination, evaluating the last factor in Eq.~\eqref{eq:inflation} leads to $\Omega_\text{gw}(f) \propto f^2$.
The GW spectrum of single field slow-roll inflation spans many orders of magnitude in frequency, corresponding the fluctuation modes exiting the horizon at different times over the course of inflation. Unfortunately, the amplitude is too small to be detectable in the foreseeable future.

However, more elaborate inflation models can lead to larger GW signals. For example, in axion inflation, an axion-like inflaton particle couples to massless Abelian gauge fields through a Chern-Simons interaction. This leads to a tachyonic instability and hence to an exponential growth of the gauge fields during inflation. The resulting gauge field energy momentum tensor constitutes a significant classical source of gravitational wave production, which can dominate over the vacuum contribution from single field slow-roll towards the end of inflation (i.e.\ at far sub-CMB scales). This could lead to an appreciable, and potentially observable, GW signal at interferometer scales~\cite{Cook:2011hg,Barnaby:2011qe}.\footnote{
The large energy densities required in the anisotropic gauge field energy to generate a sizeable GW signal entail a sizable backreaction on the background dynamics and are a challenge for perturbative computations. For example, the gauge field production results in a non-linear friction term for the axion-like particle~\cite{Domcke:2020zez} while in turn the production of any light charged fermions drains the gauge field energy~\cite{Domcke:2018eki}. Lattice simulations can accurately capture these non-perturbative effects~\cite{Figueroa:2023oxc}, however struggle to cover the relevant dynamical time scales and the full complexity of realistic models. These challenges in performing accurate predictions for strong GW signals is exemplary for many other similar examples.}

Another example are inflation models which feature an ultra-slow-roll phase and hence a very strong growth of the scalar power spectrum $\Delta_s^2(k)$ at sub-CMB scales. The combination of two such scalar fluctuations can form a tensor fluctuation, resulting in a GW spectrum which can be parametrized as~\cite{Kohri:2018awv}
\begin{align}
 \Omega_\text{gw}(f) \sim 10^{-9} \left( \frac{\Delta_s^2}{0.01} \right)^2 \frac{\Omega_\gamma}{10^{-5}} \, S(f)\,.
\end{align}
The last factor depends on the shape of the scalar power spectrum and has been normalized to have a maximum value of around unity. A detectable GW signal thus requires a very significantly enhanced scalar power spectrum, in this case an enhancement by about seven orders of magnitude compared to CMB scales to reach the current PTA sensitivity. For more details on GW signals from inflation models see e.g.~Ref.~\cite{Bartolo:2016ami} and references therein.

\paragraph{Cosmic strings.}
Cosmic strings are one-dimensional topological defects which are formed when a symmetry group $G$ is spontaneously broken to a subgroup $H$ with $\Pi_1(G/H) \neq 1$, where $\Pi_1$ denotes the first homotopy class. The simplest example is the spontaneous breaking of a $U(1)$ symmetry. Mapping the vacuum manifold of a Mexican hat potential onto 3-dimensional space allows for configurations where the symmetry is unbroken only along one dimensional `strings' which cannot be removed by infinitesimal adjustments of the symmetry breaking field. These objects are thus topologically stable.

Once formed, the cosmic strings can (self-)intersect to form a network of long strings and loops, both of which emit gravitational radiation. Once can show that the network of long strings and loops reaches and subsequently maintains a scaling regime, in which the fraction of energy density stored in the network compared to the total energy density of the universe remains constant. The GW source is thus active from the symmetry breaking process throughout the subsequent evolution of the universe, leading to a GW spectrum that spans many orders of magnitude in frequency. As for inflation, we expect a scale invariant spectrum for GWs sourced during radiation domination.

The evolution of the network and the GW emission can be modelled through numerical simulations. This has been done by classical field theory simulations of the Abelian Higgs model~\cite{Figueroa:2012kw} or by simulating the 1+1 dimensional Nambu-Goto word-line (see~\cite{Auclair:2019wcv} for a review). There is currently significant disagreement on the amplitude of the resulting GW spectrum  between the two approaches. The Nambu-Goto method predicts the larger amplitude for the plateau from radiation domination,
\begin{align}
 \Omega_\text{gw} \sim \Omega_\gamma G \mu \,,
\end{align}
with $\mu$ denoting the string tension, related to the symmetry breaking scale $v$ as $\mu \sim v^2$. Current GW experiments are probing symmetry-breaking scales around the scale of grand unification, with the most stringent constraints currently coming from pulsar timing arrays.\footnote{In some GUT embeddings can decay on cosmological time scales through the pair creation of monopoles along the string core. For these metastable strings, the GW spectrum at low frequencies is suppressed, and the most stringent bounds currently come from ground-based interferometers~\cite{Buchmuller:2021mbb}.}

\vspace{1cm}
In conclusion, while our standard models of particle physics and cosmology do not predict sizable GW signals from the early universe, a plethora of well-motivated models extending these (and addressing other open problems in cosmology) do. The search for SGWBs will thus not only yield new insights on astrophysical GW sources, but also advance our understanding of the early universe and particle physics. Key to achieving this is a robust theoretical predictions of the GW signals of different BSM models. Equally important is the data analysis challenge of understanding how a SGWB manifests in a given GW detector and how it can be disentangled from transient signals, astrophysical foregrounds and instrument noise.

\vspace{1cm}
\paragraph{Acknowledgments.} It is my pleasure to thank the TASI 2024 organizers for putting together this great school and in particular all students for their remarkable enthusiasm and their never-ending questions. Special thanks to Joydeep Naskar for comments on the manuscript. The TASI 2024 school was supported by the National Science Foundation.

\appendix

\section{Notation and Conventions}
\label{app:notaton}
We use the $(-,+,+,+)$ signature for the metric, denote 3-vectors as $\vec x$ and 4-vectors as $x$. We work in natural units with $c = \hbar = 1$.  We denote the reduced Planck mass with $M_{pl}$, related to Newtons constant as $M_{pl}^2 = 1/(8 \pi G)$.

We largely avoid the use of any explicit form of the GW tensors. A convenient definition is obtained by introducing the orthormal system spanned by the GW normal vector $\hat k$, the polar angle unit vector $\hat v = \hat e_\phi$ and $\hat u = \hat v \times \hat k$,
\begin{align}
 \hat k = \begin{pmatrix} \sin \theta \cos \phi \\ \sin \theta \sin \phi \\ \cos \theta \end{pmatrix} \,, \quad
 \hat v = \begin{pmatrix} - \sin \phi \\ \cos \phi \\ 0 \end{pmatrix} \,, \quad
\hat k = \begin{pmatrix} \cos \theta \cos \phi \\ \cos\theta \sin \phi \\ - \sin \theta \end{pmatrix} \,.
\end{align}
The polarization tensors can then be obtained as
\begin{align}
 e_{ij}^+ = u_i u_j - v_i v_j \,, \quad e_{ij}^\times = u_i v_j + v_i u_j \,.
\end{align}
By construction they are transverse  $k_i e^{ij}_\lambda(\hat k) = 0$ and orthogonal $e^\lambda_{ij}(\hat k) e^{ij}_{\lambda'}(\hat k) = 2 \delta_{\lambda \lambda'}$. The normalization is a matter of convention (many references introduce an extra $1/\sqrt{2}$ such that $e^\lambda_{ij}(\hat k) e^{ij}_{\lambda'}(\hat k) =  \delta_{\lambda \lambda'}$) and can be re-absorbed in the GW amplitudes $h^{+,\times}$. There is nothing fundamental in this particular choice of $\hat v$ and $\hat u$, in particular rotating them by $\pi/4$ around the $\hat k$ axis would equally give an orthonormal system, but would interchange what we label $+$ and $\times$.

\bibliographystyle{utphys}
\bibliography{refs}

\end{document}